# NEW ALGEBRAIC-NUMERIC METHODS FOR LOOP INTEGRALS.
# SOME 1-LOOP EXPERIENCE

D.Yu. Bardin[a], L.V. Kalinovskaya[a] and F.V. Tkachov[b]

[a] Joint Institute for Nuclear Research, Dubna, Moscow Region, 141980 Russia

[b] Institute for Nuclear Research of Russian Academy of Sciences
Moscow, 117312 Russia

We report our experiences with the generalized integration-by-parts algorithm [hep-ph/9609429] in the context of calculations of a realistic one-loop subset of diagrams.

This project combined the experience of the ZFITTER[1] team [1] with the experience of designing algorithms for multiloop calculations[2] [4], [5] in order to explore the potential of the "algebraic" scenario of loop calculations suggested in [6].

The scenario generalizes the idea of [4], which is to exploit the rich algebraic structure of integrands in order to reduce them to a maximally simple form, depending on the specific application. The original algorithm of [4] was about reduction of massless self-energies to recursively one-loop integrals easily calculable analytically; it demonstrated that such a reduction may be an extremely powerful calculational tool. The scenario of [6] is about reduction of integrands to a form maximally suitable for subsequent numerical integration. It is based on a subtle algebraic fact (a generalization of the so-called Bernstein theorem [7]) and is applicable to arbitrary loop integrals (arbitrary topologies and patterns of masses and external momenta). This is in theory.

In practice, the complexity of the problem of finding the differential identities required by the scenario of [6] for integrands with 2 or more loops now appears to be so great (and the resulting identities so hugely cumbersome) that the feasibility of the scenario of [6] beyond one loop level is rather doubtful.

However, for one-loop integrals, all algebraic difficulties reduce to an iterative application of a simple differential identity:

---

[1] ZFITTER [1], along with TOPAZ0 [2], is a program which incorporates all the available theoretical knowledge about the $Z$-resonance in the process of $e^+e^-$ annihilation [3]. They have been used to extract Standard Model parameters from the precision data obtained at LEP1 by the experimental collaborations at LEP.

[2] These algorithms are behind the flourishing industry of NNLO QCD calculations (sum rules and moments of DIS structure functions; mass expansions; $\sigma_{tot}(e^+e^- \to hadrons)$).

$$\frac{1}{\Delta}\left(1 - \frac{(x+A)\partial_x}{2(\mu+1)}\right)V^{\mu+1}(x) = V^\mu(x).\qquad 0.1$$

Here $x$ is the vector of Feynman parameters, $V(x) = x^T \tilde{V} x + 2R^T x + Z$ is a quadratic polynomial, $A = R^T \tilde{V}^{-1}$ and $\Delta = (Z - R^T \tilde{V}^{-1} R)$.

Any one-loop integral can be represented as

$$\int_S dx\, Q(x) V^{-n-\varepsilon}(x).\qquad 0.2$$

An iterative application of 0.1 followed by integration by parts reduces 0.2 to a sum of the form

$$\sum \int_{S'} dx'\, Q'(x') V^{N-\varepsilon}(x'),\qquad 0.3$$

where the sum involves integrals over simplices with no higher dimensionality than in the original integral 0.2. The Laurent expansion in $\varepsilon$ yields integrals of the form

$$\int_{S'} dx'\, Q'(x') \times V^N(x') \ln\frac{V(x')}{\mu^2},\qquad 0.4$$

where $\mu$ is the unit of mass. (In a general one loop integral second powers of log $V$ are possible.)

By choosing $N$ large enough (which is achieved by suitably many applications of the identity 0.1) one can achieve an arbitrarily large degree of smoothness of the integrands in 0.4, ensuring their better amenability to numerical integrations. With $N=0$, one has a simple absolute convergence (logarithmic singularities in the integrand). With $N=1$, the integrand is continuous; with $N=2$, it has continuous first derivatives, etc. Simple 1-dimensional examples show that the optimal rate of numerical convergence (with integration algorithms that make use of continuity of derivatives) is achieved for values of $N=4,5$; by rate of convergence we understand the number of integrand evaluations required to attain a precision of 8-10 digits.

On the downside, the quantity $\Delta$ which appears in the denominator may contain small factors near thresholds, leading to large numerical cancellations between different terms in the sum 0.3, which offset the benefits of smoother integrands.[3] The scenario of [6] offers no options for that. Note, however, that a complete set of rules for obtaining systematic asymptotic expansions of arbitrary Feynman integrals in arbitrary asymptotic regimes was obtained in [8], so the handling of cases with a small $\Delta$ is in theory not an insurmountable problem (although, of course, one would prefer a uniform algorithm to treat the various cases).

***Our aim*** was to explore the behavior of this algorithm in a realistic case of a subset of diagrams which contribute to the ZFITTER [1]. The subset (which we call $Z$ cluster; for details see [9]) is gauge invariant and UV finite. It is shown in the following figure:

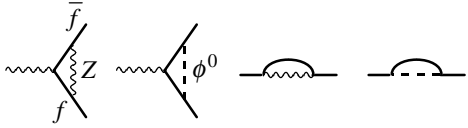

We will consider and evaluate three standard scalar formfactors $F_\#^{\gamma Z}$, $\# = L, Q, D$ which are present in the $Z$ cluster with incoming photon $\gamma$ in the $Z$ vertex. These can be expressed in terms of the standard Passarino-Veltman functions $C_0$ and $B_0$ (see e.g. [3]). For instance:

$$i\pi^2 C_0(p_1^2, p_2^2, Q^2; m_1, m, m_2) = \mu^{n-4} \int d^n q \, \frac{1}{d_0 d_1 d_2}, \quad 0.5$$

where $d_0 = q^2 + m_1^2 - i\varepsilon$, $d_1 = (q+p_1)^2 + m^2 - i\varepsilon$ and $d_2 = (q+Q)^2 + m_2^2 - i\varepsilon$. Also $Q = p_1 + p_2$. In terms of Feynman parameters (in $n=4$):

$$C_0(p_1^2, p_2^2, Q^2; m_1, m, m_2) = \int_0^1 dx \int_0^{1-x} dy \, P_{x,y}^{-1}, \quad 0.6$$

where

$$P_{x,y} = -x^2 p_1^2 - y^2 p_2^2 + xy(Q^2 - p_1^2 - p_2^2)$$
$$+ x(p_1^2 + m_1^2 - m^2) + y(p_2^2 + m_2^2 - m^2) + m^2. \quad 0.7$$

We introduce the following generalizations:

$$C_0^{(k)}(\ldots) = \int_0^1 dx \int_0^{1-x} dy \, P_{x,y}^k \ln \frac{P_{x,y}}{\mu^2}. \quad 0.8$$

Similarly:

$$B_0^{(k)}(\ldots) = \int_0^1 dx \, P_x^k \ln \frac{P_x}{\mu^2}, \quad 0.9$$

where

$$P_x = Q^2 x(1-x) + M_1^2 x + M_2^2 (1-x). \quad 0.10$$

Conventional expressions for the formfactors $F_\#^{\gamma Z}$, $\# = L, Q, D$ are in terms of the standard $C_0$ and $B_0$.

One application of the identity 0.1 (in $n \neq 4$) together with integration by parts yields expressions in terms of $C_0^{(0)}$ and $B_0^{(0)}$. The total length of the three expressions is one LaTeX page. Two applications yield expressions in terms of $C_0^{(1)}$ and $B_0^{(1)}$, with the total length of two pages.

After the algebraic simplifications, numerical integration was performed using a simple Simpson integrator which is sensitive to smoothness of integrands. The numerical results are presented in the following table:

|  | $k$ | Re | Im |
|---|---|---|---|
| $F_L^{\gamma Z}$ |  | 1.62989336 | 2.32844153 |
|  | 0 | $<10^{-9}$ | 1.5E-3 |
|  | 1 | $<10^{-9}$ | 1.9E-5 |
| $F_Q^{\gamma Z}$ |  | -0.366725194 | 0.808473633 |
|  | 0 | $<10^{-9}$ | 2.0E-4 |
|  | 1 | $<10^{-9}$ | 2.2E-6 |
| $F_D^{\gamma Z}$ |  | -0.00234117123 | -0.0063346893 |
|  | 0 | $<10^{-9}$ | 1.9E-3 |
|  | 1 | $<10^{-9}$ | 2.3E-5 |

For each formfactor, the first line lists exact results (obtained in the usual way via dilogarithms). The second and third lines give absolute relative errors for the results of numerical calculation of the same quantities. The lines $k=0|1$ correspond to the cases when the identity 0.1 is used once|twice so that all resulting integrands have the singularity $\ln P | P \ln P$ (absolute integrability with logarithmic singularities [step function in the imaginary parts] in the first case; continuous integrands in the second case). The simplest Simpson integrator was used which likes continuous second order derivatives. But we felt the complexity of the formulas resulting from three application of the differential identity was too great to justify the effort involved in the interfacing of the algebraic and numerical parts of this project for more than two applications of the identity 0.1 at the current exploratory stage of the project.

---

[3] Examples of this kind were provided to us by K. Kato. Note that maximal benefits from the algebraic transformations being described are obtained with integration routines that make use of the increased regularity of resulting integrands. But since the method effectively subtracts *all* types of divergencies/singularities simultaneously, it may be beneficial even with MC integrators.

*A conclusion* that can be drawn from our experiment is that the scenario based on the use of the algebraic identity 0.1 may, at least in some one loop applications, be a viable alternative to the standard procedures based on the use of analytical expressions in terms of dilogarithms (cf. [3]). Moreover, we have observed a number of attractive properties:

- Gauge cancellations within gauge invariant subsets of diagrams occur prior to numerical integrations. This means that numerical cancellations due to at least this source are not encountered.

- Similarly cancel UV divergences.

- In our example, we have observed little (if any) evidence of numerical instabilities. It is essential here that a deterministic numerical integration algorithm was used which took into account the increased smoothness of the integrands; Monte Carlo integration algorithms are inappropriate here.

- The method offers an interesting alternative to the conventional Passarino-Veltman reduction to scalar integrals. Note that the new reduction (i.e. the representation in terms of the basic scalar functions 0.8, 0.9 with all $x'$ excluded from 0.4) was implemented in $n$ dimensions.

*The greatest technical difficulty* for implementation of the described scenario — both in one loop and to a very much greater extent beyond one loop — appears to be the dominant software engineering platform in the HEP community (conventional inefficient symbolic manipulation programs + numerical calculations with the archaic Fortran/C or the clay-feet monster of C++, with no easy connection between the two). Another problem is that the computer resources needed even in simpler cases are huge (although Moore's Law is slowly taking care of that; perhaps too slowly).

Since there are physical problems in which further progress via conventional approaches seems to be all but excluded, experimentation with the scenario for loop calculations of [6] will continue. On a limited scale and in some special problems one may even expect realistic calculations to occur within a few years — provided a sufficient high-quality software engineering effort could be mustered. But the main problem is the inadequacy of the dominant software engineering platform in HEP.[4]

*Acknowledgments*

This work was supported in part by the Russian Foundation for Basic Research (quantum field theory section) under grant 99-02-18365. F.T. thanks K.Kato for providing some numerical examples of calculations of one-loop integrals.

---

[4] It cannot be emphasized too strongly: C++ cannot be a foundation on which to build future. Some related comments together with a discussion of alternatives are collected in [10].